\documentclass[aip,jap,amsmath,amssymb,reprint,floatfix]{revtex4-2}

\usepackage{graphicx}
\usepackage{dcolumn}
\usepackage{bm}
\usepackage[utf8]{inputenc}
\usepackage[T1]{fontenc}
\usepackage{upgreek}
\usepackage[normalem]{ulem}
\usepackage{hyperref}
\hypersetup{
    colorlinks=true,
    linkcolor=blue,
    filecolor=blue,      
    urlcolor=blue,
    citecolor=blue,
}

\begin{document}

\title{Bright Spatially Coherent Beam from Carbon Nanotube Fiber Field Emission Cathode}
\author{Taha Y. Posos}
\email{posostah@msu.edu}
\affiliation{Electrical and Computer Engineering, Michigan State University, MI 48824}



\author{Sergey V. Baryshev}
\email{serbar@msu.edu}
\affiliation{Electrical and Computer Engineering, Michigan State University, MI 48824}

\begin{abstract}
 Large area carbon nanotube (CNT) cathodes made from yarns, films or fibers have long been promising as next generation electron sources for high power radio frequency (rf) and microwave vacuum electronic devices. However, experimental evidence have been pointing out spatial incoherence of the electron beam produced by such cathodes that, in turn, impeded the progress toward high brightness CNT electron sources and their practical applications. Indeed, typically large area CNT fibers, films or textiles emit stochastically across their physical surface at large emission angles and with large transverse spread, meaning large emittance and hence low brightness. In this work, using high resolution field emission microscopy, we demonstrate that conventional electroplating of hair-thick CNT fibers followed by a femtosecond laser cutting, producing emitter surface, solves the described incoherent emission issues extremely well. Strikingly, it was observed that the entire (within the error margin) cathode surface of a radius of approximately $75\,\mu\text{m}$ emitted uniformly (with no hot spots) in the direction of the applied electric field. The normalized emittance on the fiber surface was estimated of 52 nm with brightness of $>$$10^{15}\frac{\text{A}}{\text{m}^2\text{rad}^2}$ (or $>$$10^7$ A m$^{-2}$sr$^{-1}$V$^{-1}$) estimated for pulsed mode operation.
\end{abstract}


\maketitle

\section{\label{S:Intro}Introduction}
In early 1990's, carbon nanotechnology revolution introduced a plethora of new advanced materials among which the carbon nanotube was notoriously attractive for making nanoscale field effects devices, including vacuum devices. Many labs studied effects associated with field emission from a single CNT or arrays with counted number of isolated CNTs.\cite{Motorola1, Thales, BrightCNT, HotCNT_PRL, CNT_arrays_apps1, CNT_arrays_apps2} Control over fabrication and emission of single CNT field emission devices was excellent and many field emission devices were demonstrated, e.g. field emission radio\cite{CNTradio} or field emission transistor\cite{CNT_FE_transistor}, amplifier\cite{CNT_FE_amp}, and many others.\cite{CNT_arrays_apps1,CNT_arrays_apps2}

\begin{figure}
\includegraphics[width=5.5cm]{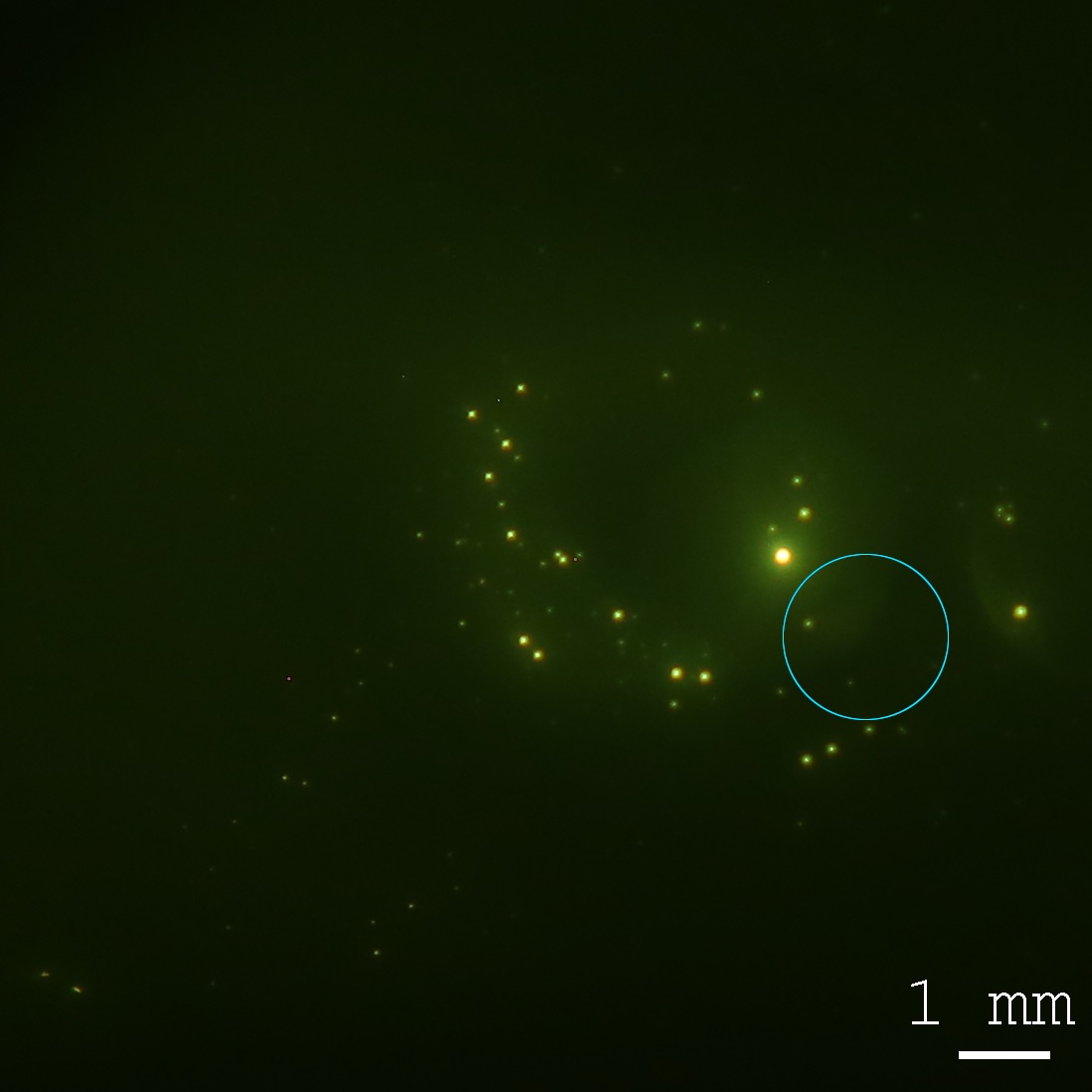}
\caption{Typical micrograph showing large beam transverse spread and nonuniformity. The blue circle marks the cathode's position behind the imaging screen and its size. }\label{F:nonuniform}
\end{figure}

In order to increase the output power, macrosocopic large area CNT fibers, films, yarns and fabrics started to be used to increase the operating currents from pico-/nano- to many amperes. Here, CNTs were thought to replace legacy velvets.\cite{CNTvsVelvet} The multiple benefits of CNT fibers over legacy technology are low turn-on voltage and high emission current at relatively low operating electric field due to inherent high field enhancement factor, and high electrical and thermal conductivities.\cite{fabrication} It was conventionally assumed that emission would be uniform, i.e. uniformity would translate from previously studied arrays of counted CNTs to the large area CNT fibers. However, recent studies that employed field emission microscopy illustrated that emission is never uniform and moreover that the emission area is a function of the electric field (making it cumbersome for calculating current densities.) Fig.\ref{F:nonuniform}, reproduced from our past work,\cite{mypaper} highlights other important issue of the large transverse spread of the emitted beam where beam lands on the imaging screen millimeters away from the physical location of the cathode source (blue circle) after travelling only a millimeter between the cathode and the anode. This clearly points out a very large emittance and therefore very low brightness, making CNT fiber cathodes impractical for applications like rf or microwave traveling wave tubes (operating in GHz range), microscopy and bright X-ray sources for medicine or active scanning. Another issue arises from that---because all the current emerges from a few active spots, it leads to local heating, microbreakdowns\cite{mypaper} and short-lived cathodes.

After experimenting with many fiber arrangements, we found that it is tiny singular fibrils (comprising braided fibers) that set loose due to thermal and field related stress\cite{mypaper,morphology_dependent} and that eventually focus the field due to their high aspect ratio and become point-like randomized intense electron emitters eventually exploding (seen as micro-breakdowns) and re-populating surrounding areas with more new-born fibrils. This process repeats itself until the cathodes stops operating while emission always look like a family of single electron rays going in many directions that are not aligned with the desired main longitudinal propagation direction, such as in Fig.\ref{F:nonuniform}. To mitigate this issue, hypothesised to be the major problem, we study new cathode production technology where fibers are electroplated with Ni and laser cut; all to suppress the fibril occurrence and regeneration. Through experimental measurements and electrostatic and beam dynamics modeling, emission uniformity and beam brightness were analyzed.

\section{\label{S:experimental}Experimental}
\begin{figure}
    \centering
    \includegraphics[scale=0.045]{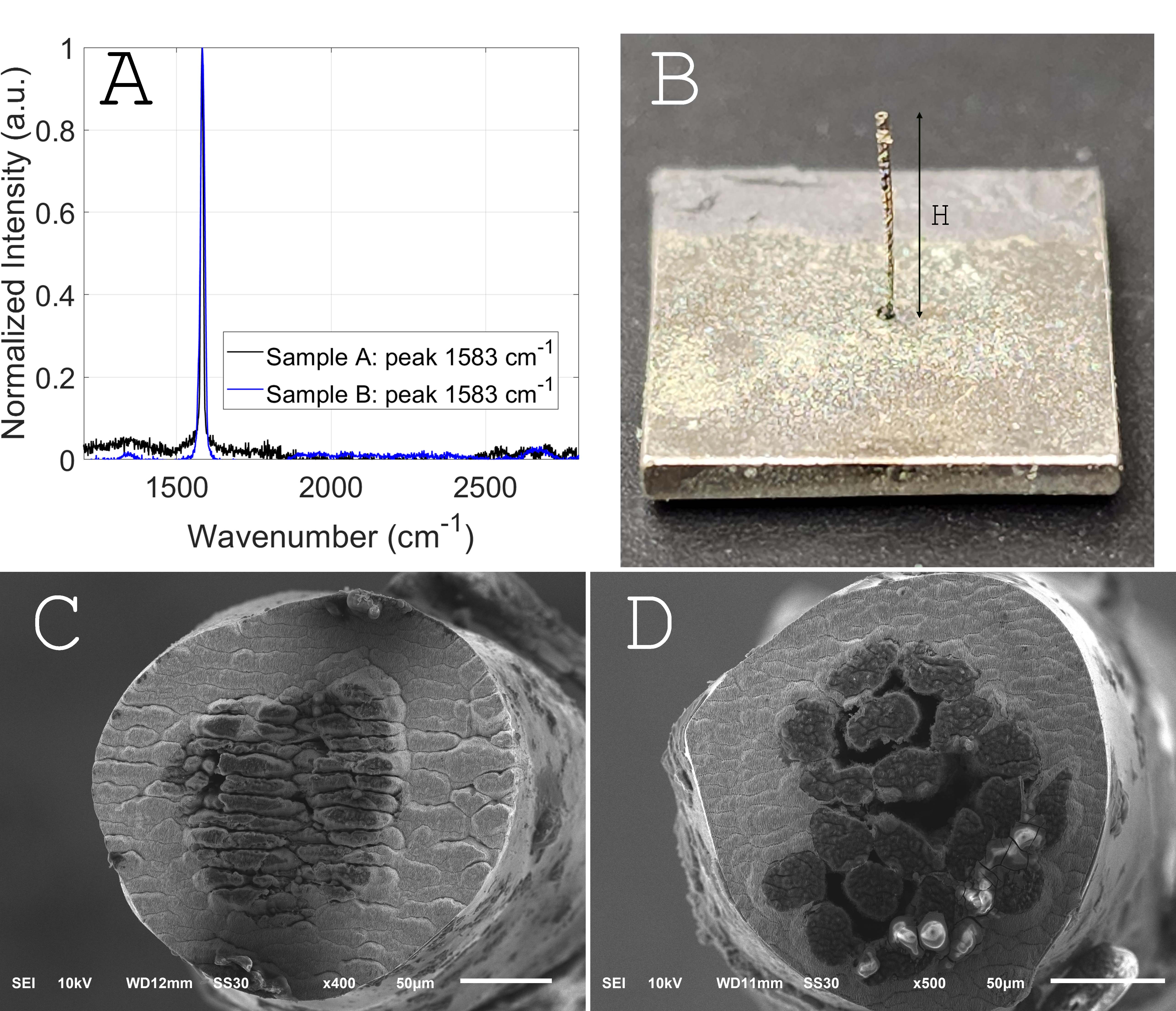}
    \caption{(A) Raman spectra of the cathode surface showing a crystalline graphitic peak. (B) Electroplated CNT fiber welded on a Ni base; $\text{H}=4.8\,\text{mm}$ for Sample A and $\text{H}=4.6\,\text{mm}$ for Sample B. (C) and (D) SEM images of Sample A and Sample B, where scale bars are 50 $\mu$m.}
    \label{F:sample}
\end{figure}

To prepare the field emission CNT fiber cathodes, a commercially available CNTs fiber from DexMat, Inc was used. The fiber is made by a wet-spinning technology \cite{fabrication}---pre-grown arrays of CNTs are dissolved in an acid to form a spinnable liquid dope that is extruded through a spinneret into coagulant bath to remove acid, and then dried in an oven. The resulting product is highly aligned and densely packaged CNTs in a form of a fiber. DexMat fibers have high electrical and thermal conductivity. Such fibers were shown to feature anisotropic field emission\cite{unisotropic_emission} that is emission takes place along the fiber (not from side walls) which is a great property allowing for control over emittance. Raman spectroscopy shows the G peak positions at 1583 cm$^{-1}$ suggesting rich crystalline graphitic content as expected from high quality fibers (Fig.\ref{F:sample}A).

To mitigate the described stray fibril problem, few fibers of the described kind were placed side by side and electroplated with Ni in an electrochemical bath and flush cut from the top to the required length of about $5\,\text{mm}$ with a femtosecond micromachining laser beam. Then, it was welded on $1\times 1\,\text{inch}$ Ni base. The final fabricated structure can be seen in Fig.\ref{F:sample}B.

We tested two samples referred to as Sample A and Sample B through the rest of the paper. In Sample A fibers were twisted and in Sample B were not, i.e. simply place along each others. As scanning electron microscopy (SEM) demonstrates in Fig.\ref{F:sample}C, additional fiber twisting enabled a dense core in Sample A, while Sample B (Fig.\ref{F:sample}D) has visible voids between the individual fibers. Otherwise, both sample have fiber core diameter of $\sim 150\,\mu\text{m}$ and Ni shell thickness of $\sim50\,\mu\text{m}$. Sample A and Sample B have height of $4.8\,\text{mm}$ and $4.6\,\text{mm}$ respectively (see Fig.\ref{F:sample}B).


DC current tests and field emission microscopy were performed in our custom field emission microscope described in great detail in Ref.\onlinecite{exp_setup}. Images were processed by a custom image processing algorithm FEpic described elsewhere.\cite{mypaper2}

\section{\label{S:ivrelation}Field Emission Imaging and Conditioning}

\begin{figure}
\includegraphics[scale=0.47]{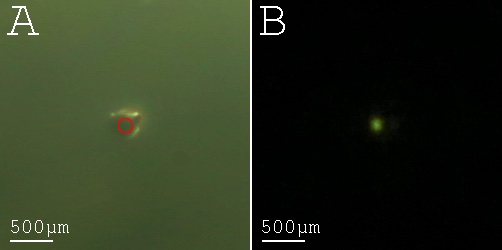}
\caption{A) Sample A seen through the YAG screen when the lights is on in the chamber. Its fiber core is marked with red circle. B) FE micrograph of the same region at the gap of $200\,\mu\text{m}$.}\label{F:sampleA}
\end{figure}

\begin{figure}
\includegraphics[scale=0.47]{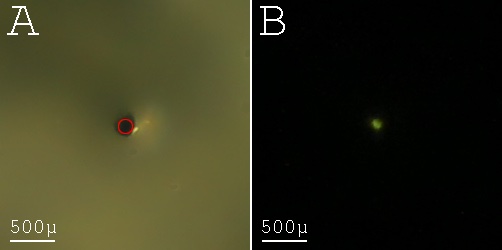}
\caption{A) Sample B seen through the YAG screen when the lights is on in the chamber. Its fiber core is marked with red circle. B) FE micrograph of the same region at the gap of $200\,\mu\text{m}$.}\label{F:sampleB}
\end{figure}

After sample was installed and gap was tuned using a doublet of two orthogonal optical microscopes, the physical location of the fiber is determined and labeled. To do that, the test chamber is illuminated. Because the imaging anode YAG screen is semitransparent, the location of the fiber can be immediately seen and captured by photographing. The core of the fiber is marked with red circle for reference in Figs.\ref{F:sampleA}A and \ref{F:sampleB}A.

After that voltage is applied and field emission images are taken concurrently with I-V curves. Fig.\ref{F:sampleA}B shows emission micrograph of the Sample A. The improvement is immediately obvious when compared with Fig.\ref{F:nonuniform}. First, the emission spot appears exactly at the optical projection of the cathode. This means beam divergence angle is small, so emittance can be expected to be low. Second, there is only a single spot and its size is comparable to the size of the fiber core---this is an indication of uniformity and small angular spread of the electron beam. The same exact behaviour was observed for Sample B, as given in Fig.\ref{F:sampleB}. No evidence suggesting the stray fibril issue was observed for neither cathode. These results highlight that such a simple electroplating strategy is extremely effective at yielding emission uniformity and spatial coherence, thereby boosting the transverse beam brightness. 

\begin{figure}
\includegraphics[width=7.cm]{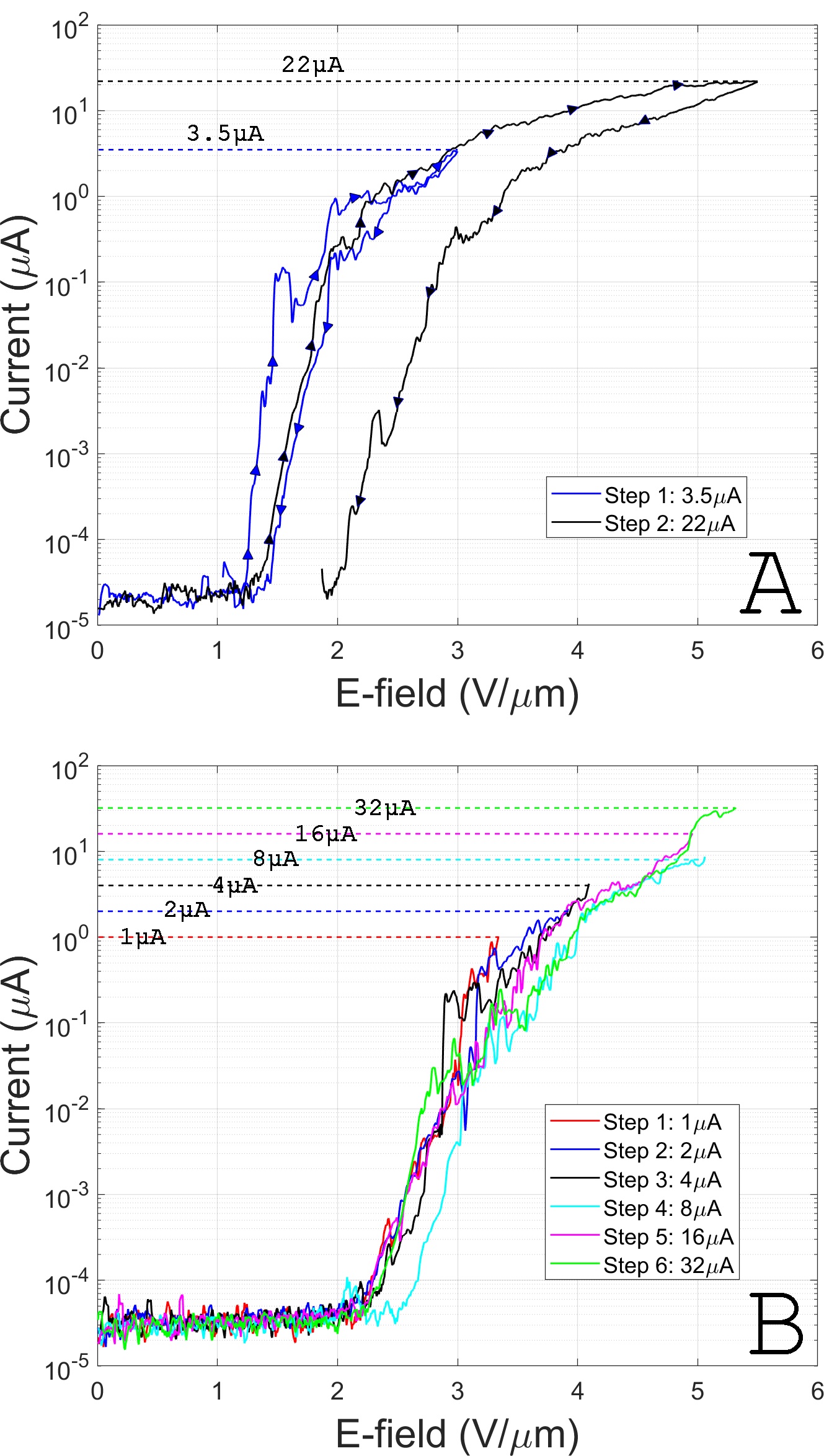}
\caption{A) The conditioning scheme of Sample A. Both ramp up and down curves are shown. There is a clear decrease in performance. B) Conditioning scheme of Sample B. Only ramp-up curves are shown. There is no considerable change in performance.}\label{F:condition}
\end{figure}


Fig.\ref{F:condition} shows cathode conditioning I-V curves for Sample A and B at interelectrode gap of 200 $\mu$m. Conditioning\cite{tungsten_conditioning,diamond_conditioning,mypaper} or cycling, where the applied voltage is ramped up and down to a progressively higher number in every consecutive cycle until the desired operating current is achieved, is a crucial procedure to maximize field emission cathode performance and ensure stability at the operating point.


We found that the electroplated fibers should be conditioned with small incremental steps to avoid adverse effects such as sudden burn-down. Fig.\ref{F:condition}A demonstrates the case where the maximal field was doubled with respect to the previous conditioning cycle, such that the current went up from 3 to 22 $\mu$A. Next, after completing the ramp down Sample A stopped working completely, which was possibly due to applying electric power that exceed that in the previous run by more than an order of magnitude. Its operation could not be rejuvenated by applying higher electric fields. This is unlike a conditioning scheme that was used for Cathode B as shown in Fig.\ref{F:condition}B. The emission current was doubled at every conditioning cycle: up to 1 $\mu$A and down to 0, then to 2 $\mu$A, to 4 $\mu$A, to 8 $\mu$A, to 16 $\mu$A, and finally to 32 $\mu$A. By doing so, Cathode B was conditioned $softly$ (compared to Sample A) maintaining and enhancing its performance: $i$) the resulting operating field went up and doubled, reaching same exact value where Cathode A burned down; $ii$) turn-on field and field enhancement factor remained nearly the same meaning that Cathode B was conditioned to stably sustain higher local field.

When it is compared to our past cathode designs, detailed in Ref.\onlinecite{mypaper}, they emit less at any given field. This is an expected result because (with stray fibrils mitigated) the field enhancement is reduced. However, turn-on fields are still very low, between 1 and 2.5 V/$\mu$m. Because the beam was tight suggesting high current density we limited our measurements to between 10--100 $\mu$A as the power density deposition at the imaging screen could attain above 1 kW/cm$^2$ at the voltage source limit of 1100 V, thereby literally drilling holes in it.\cite{power-density-1, power-density-2} At 1100 V, Sample A maxed out at 20 $\mu$A and Sample B 30 $\mu$A, respectively. Again both cathodes had similar metrics. Having these metrics and qualitative results in mind, a step was taken to carry out more quantitative analysis and calculate cathodes' emittance and brightness. All detailed in the next section.


\section{\label{S:emittance}Emittance and Brightness}



In the phase space $(x,x')$, $x$ is spatial position and $x'=\frac{dx}{dz}=\frac{dx/dt}{dz/dt}=\frac{v_x}{v_z}$ is the slope of the trajectory from longitudinal centrosymmetric axis of each particle. Then, rms emittance $\tilde{\epsilon}_x$ is defined as
\begin{equation}\label{E:rmsemit}
    \tilde{\epsilon}_x = \sqrt{\langle \Delta x^2 \rangle \langle \Delta x'^2 \rangle - \langle \Delta x \Delta x' \rangle}
\end{equation}
where $\Delta x=x-\langle x\rangle$ and $\Delta x'=x'-\langle x'\rangle$. For a beam with cylindrical symmetry in $(x,y)$ and $(x',y')$ centered around zero, $\langle x\rangle$ and $\langle x'\rangle$ are zero. The cross-corelation term, $\langle \Delta x \Delta x' \rangle$ can be removed with proper beam optics.\cite{jarvis_thesis} Then, Eq.\ref{E:rmsemit} reduces to
\begin{equation}\label{E:rmsemit2}
    \tilde{\epsilon}_x = \sqrt{\langle x^2 \rangle \langle x'^2 \rangle}=\sigma_x\sqrt{\langle x'^2 \rangle}.
\end{equation}

Rms emittance is a function of the beam energy as $x'$ is changing under acceleration, and is not useful while comparing beam or beams at different energies. On the other hand, from Liouville's theorem, normalized emittance is a conserved quantity under acceleration as long as the beam is only subjected to conservative forces. The relation between rms emittance and normalized emittance is given by
\begin{equation}\label{E:normemit}
    \epsilon_x^\text{N} = \gamma\beta\tilde{\epsilon}_x
\end{equation}
where $\gamma = \frac{1}{\sqrt{1-\beta^2}}$ is the Lorentz factor and $\beta = \frac{v}{\text{c}} \approx \frac{v_z}{\text{c}}$. In our case, $\gamma\approx1$ because energy is $<=$1 keV. Mean-transverse energy, $\text{MTE}$, is $\frac{1}{2}\text{m}_\text{e} \langle v^2 \rangle$, where $v^2=v_x^2+v_y^2$. Because of the cylindrical symmetry in $(v_x,v_y)$, $\text{MTE}\approx\frac{1}{2}\text{m}_\text{e} \langle 2v_x^2 \rangle=\text{m}_\text{e}\langle v_x^2\rangle$. Then, after substituting Eq.\ref{E:rmsemit2}, in terms of $\text{MTE}$, Eq.\ref{E:normemit} becomes
\begin{equation}\label{E:fundametal}
    \epsilon_x^\text{N} = \frac{v_z}{\text{c}}\cdot\sigma_x\cdot\sqrt{\langle \frac{v_x^2}{v_z^2} \rangle}=\sigma_x\cdot\sqrt{\frac{\text{MTE}}{\text{m}_\text{e}\text{c}^2}}
\end{equation}
Practically, the normalized emittance at the cathode surface is calculated as follows. If the radius of the uniformly emitting surface of the cathode is $r_\text{i}$, $\sigma_x\approx r_\text{i}$ can be taken. Moreover, at the surface, MTE is due to statistical distribution of electrons inside the cathode itself. So, it is intrinsic, and is further redefined as $\text{MTE}_\text{i}$. Then, Eq.\ref{E:fundametal} becomes
\begin{equation}\label{E:ex_finalform}
    \epsilon_x^\text{N}=r_\text{i}\cdot\sqrt{\frac{\text{MTE}_\text{i}}{\text{m}_\text{e}\text{c}^2}}
\end{equation}

By using normalized emittance, normalized transverse brightness, $B_\text{N}$, can be calculated as
\begin{equation}\label{E:brightness}
 B_\text{N}= \frac{2\,I}{\epsilon_x^\text{N} \epsilon_y^\text{N}}
\end{equation}
where $I$ is the emitted current. $\epsilon_x^\text{N}=\epsilon_y^\text{N}$ can be taken in cylindrical symmetry.

\begin{figure}
\includegraphics[scale=0.24]{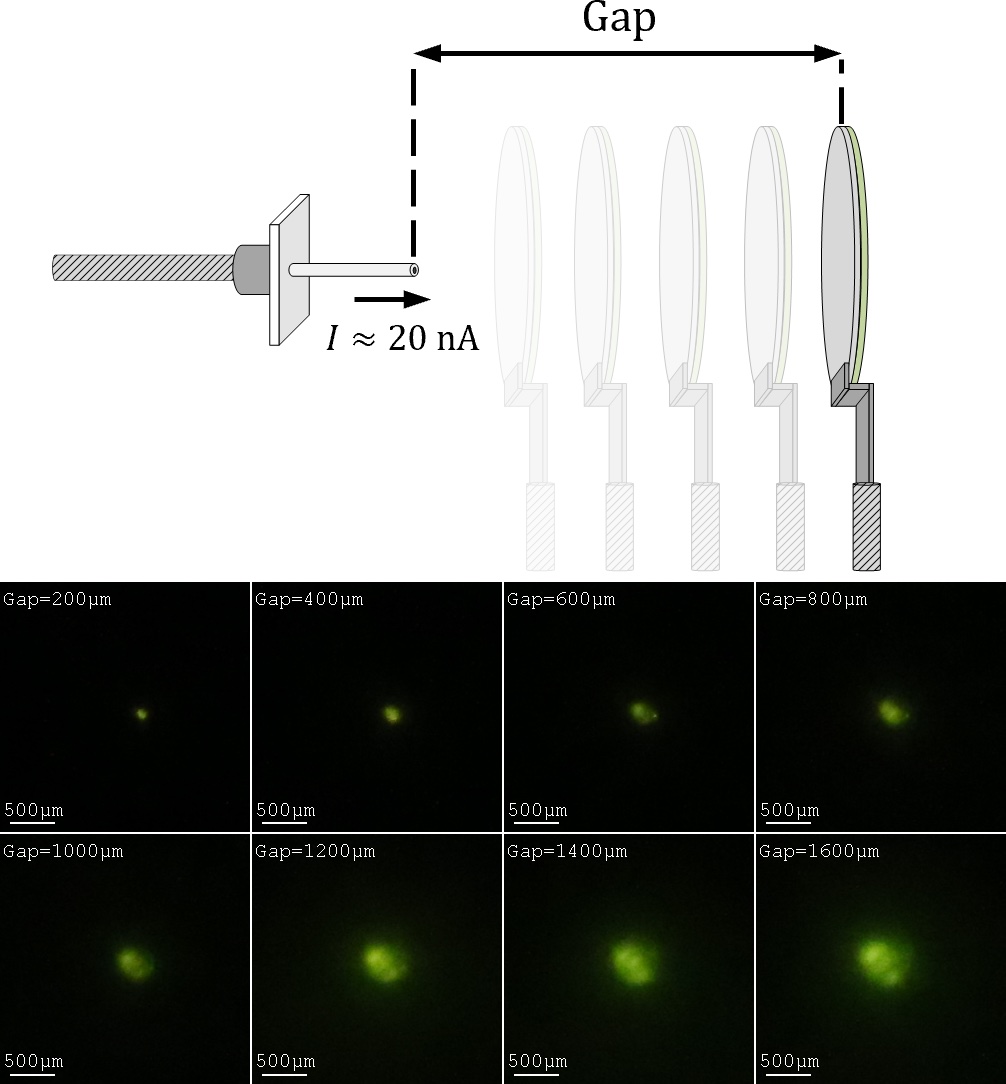}
\caption{Series of field emission micrographs of Sample B with the screen moved away progressively from $200\,\mu\text{m}$ gap to $1600\,\mu\text{m}$ gap with $200\,\mu\text{m}$ steps. At each step, a micrograph is captured.}\label{F:moveaway}
\end{figure}

\begin{figure}
\includegraphics[scale=0.2]{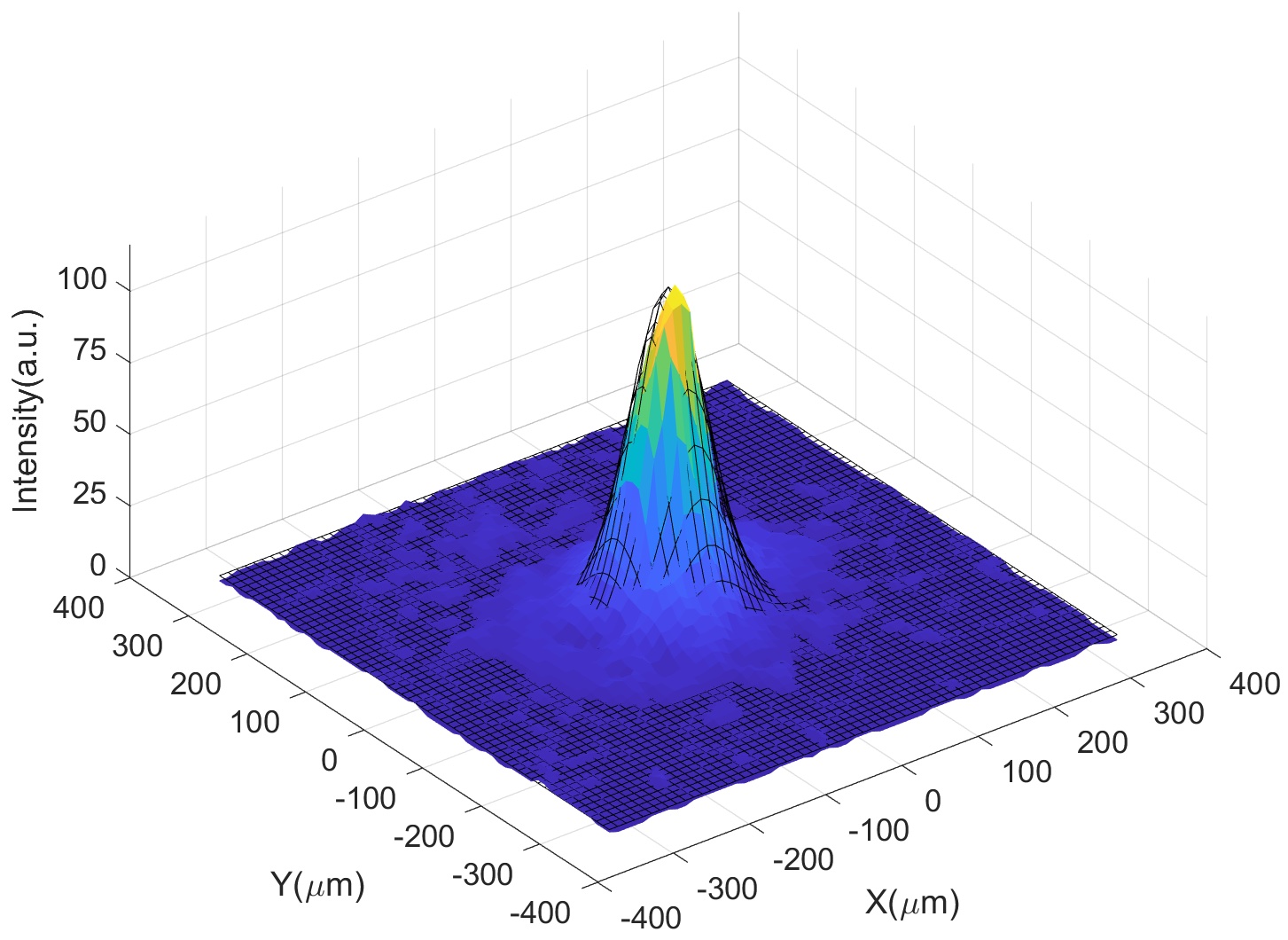}
\caption{The color surface shows the beam spot for $200\,\mu\text{m}$ gap in 3D. The black mesh surface shows its mathematical fit in Eq.\ref{E:gauss_fit}.}\label{F:gauss}
\end{figure}

\begin{figure}
\includegraphics[width=6.5cm]{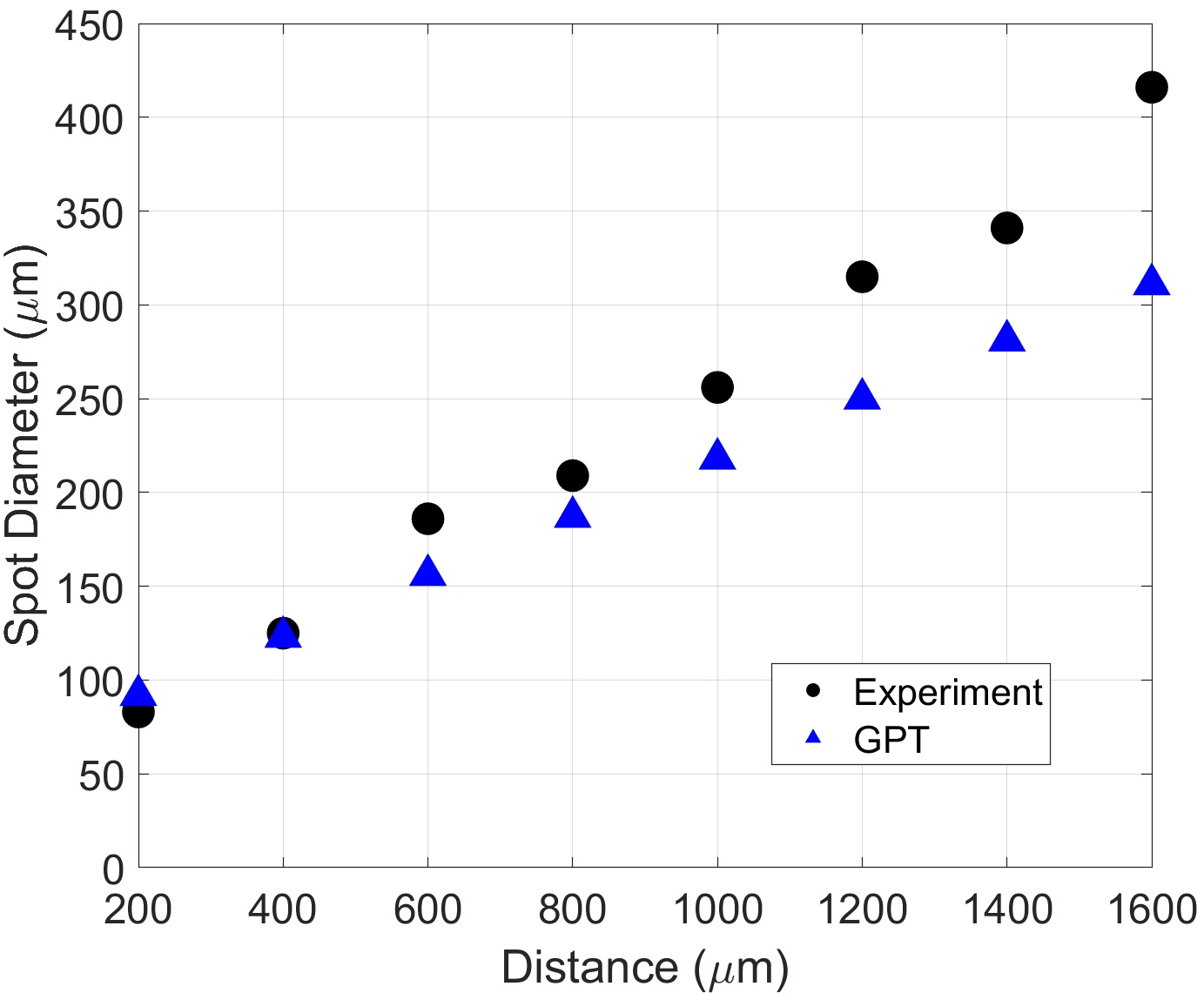}
\caption{Comparison of experimental and computational final beam spot size as the screen is moving away from the cathode. $\text{MTE}_\text{i}$ of $250\,\text{eV}$ and $r_\text{i}$ of $75\,\mu\text{m}$ were used in GPT modeling.} \label{F:gapvsize}
\end{figure}

To determine $r_\text{i}$ and $\text{MTE}_\text{i}$ in Eq.\ref{E:ex_finalform}, we conducted experimental measurements in combination with beam dynamics in GPT (General Particle Tracer).\cite{gpt_ref} In the measurements, the imaging screen is moved away from the cathode progressively. The voltage is set accordingly to keep the current constant at 20 nA to enable strong beam image signal but avoid additional beam expansion due to vacuum space charge effect. A micrograph at each step is recorded  (see Fig.\ref{F:moveaway}). Increase in the spot size due to larger time of flight is measured. As it is seen in Fig.\ref{F:moveaway}, the spots are Gaussian in nature with dense center and faint tails. Each spot can be modeled mathematically with a cylindrically symmetric Gaussian as
\begin{equation}\label{E:gauss_fit}
    p=A\cdot \text{exp}\left( -\frac{(x-x_c)^2+(y-y_c)^2}{2\sigma_{\text{spot}}^2}\right)+C
\end{equation}
to extract projected transverse beam size. Here, $A$ is the amplitude, $\sigma_\text{spot}$ is the standard deviation, $C$ is the background offset, $p$ is the intensity, $(x,y)$ are the space dimensions, $(x_c,y_c)$ are the coordinates of the peak.\cite{mypaper2} The model parameters $A$, $\sigma_\text{spot}$, and $C$ for each spot are computed with least-square fitting method. After fitting, the emission spot diameter is taken as $2\sigma_\text{spot}$. An exemplary 3D fitting done by FEpic for the beam imaged at 200 $\mu$m gap is presented in Fig.\ref{F:gauss}; with the black mesh surface being the fitting surface. The resulting dependence of the spot size, first measured (Fig.\ref{F:moveaway}) and processed by FEpic, versus distance is shown in Fig.\ref{F:gapvsize} with black solid circles. The data in the figure is only for Sample B. Because Sample A burned down, studies similar to those presented in Fig.\ref{F:moveaway} could not be carried out.

\begin{figure}
\includegraphics[width=6.5cm]{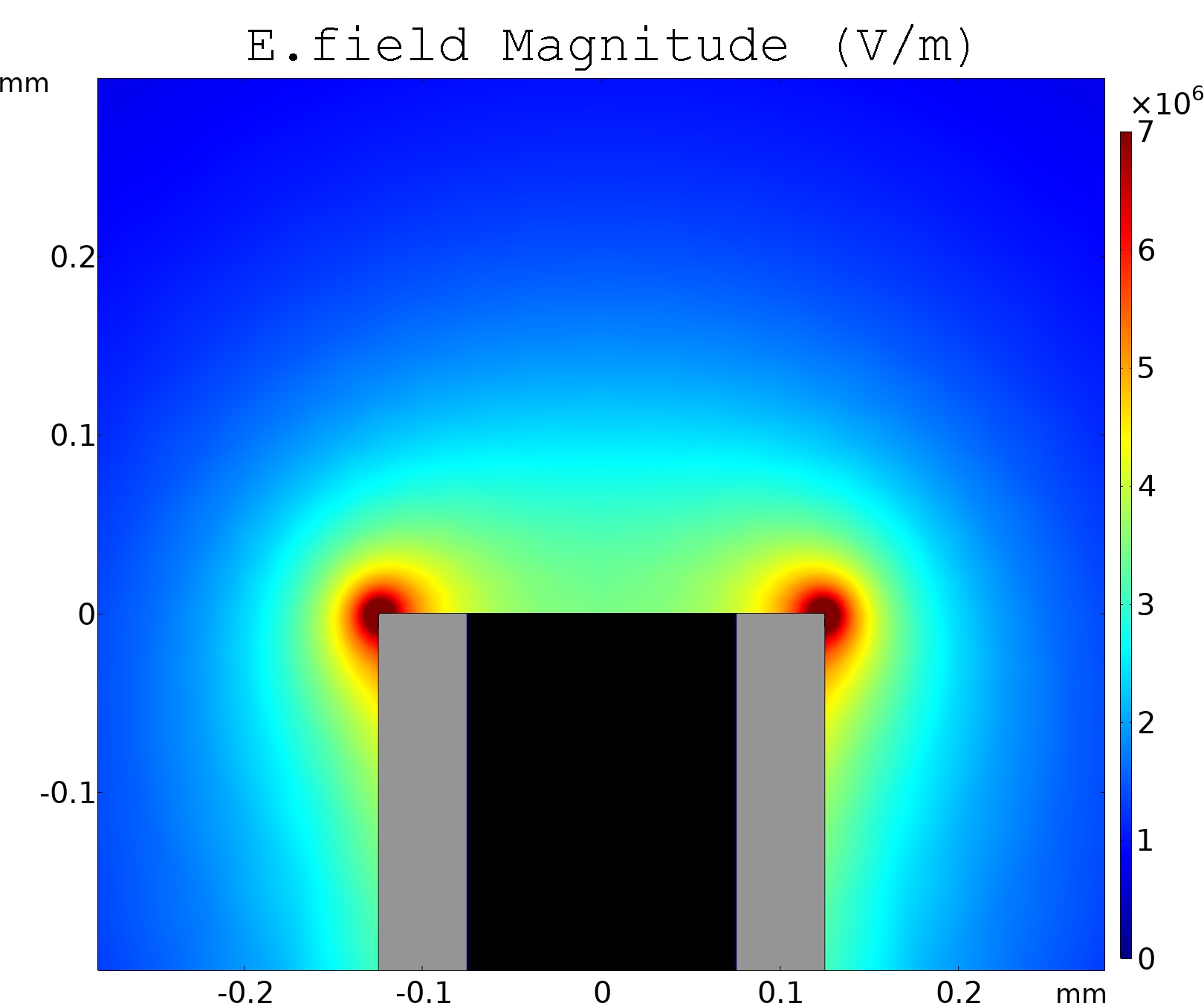}
\caption{Illustration of electric field computed in COMSOL for a $1\,\text{mm}$ gap. The color plot shows the field magnitude and contour. The dark region is the fiber core, and the gray region is the Ni shell.}\label{F:comsol}
\end{figure}

\begin{figure}
\includegraphics[width=8.5cm]{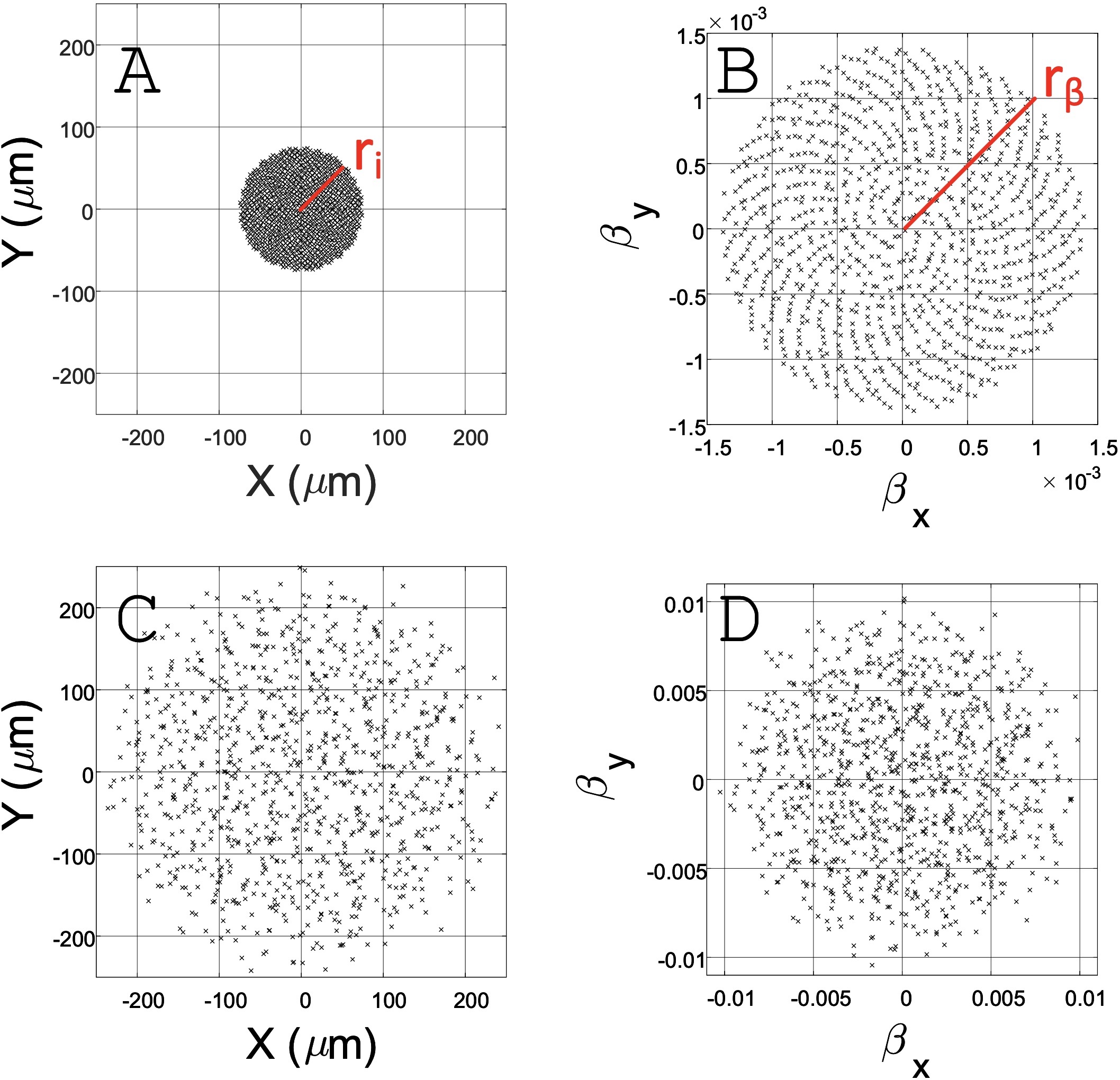}
\caption{In GPT: (A) Initial uniform beam distribution at the cathode surface in real-space, where $r_\text{i}$ is the radius of the beam. (B) The initial distribution in momentum-space, where $\beta_x=v_x/c$, $\beta_y=v_y/c$, and $r_\beta$ is the radius. Final distribution in (C) real space and (D) momentum space when the screen at is $1\,\text{mm}$.}\label{F:gpt}
\end{figure}

To calculate the phase space volume for emittance and brightness estimations and obtain $r_\text{i}$ and $\text{MTE}_\text{i}$, we switched to beam dynamics in GPT by comparing it with the experiment in Fig.\ref{F:moveaway}. To do that, a field map for each interelectrode gap was computed in COMSOL by solving Poisson's equation with given boundary conditions. An exemplary field distribution for 1 mm gap is shown in Fig.\ref{F:comsol}. Then, the field maps were imported to GPT. In GPT, the initial particle distribution in the real space $(x,y)$ (Fig.\ref{F:gpt}A)  and momentum space $(\beta_x,\beta_y)=(v_x/\text{c},v_y/\text{c})$ (Fig.\ref{F:gpt}B) has to be set to propagate the beam. Because the cathode itself circular, we used a circular uniform distribution with the radius $r_\text{i}$ in position-space and the radius $r_\beta$ in $\beta$-space. $\text{MTE}_\text{i}$ of the fiber is expected to be $250\,\text{meV}$.\cite{mte_fairchild} To be used in GPT, $\text{MTE}_\text{i}$ is converted into $r_\beta$ as
\begin{equation}\label{E:mte_vs_maxB}
    r_\beta=\sqrt{\frac{4\,\text{MTE}_\text{i}}{\text{m}_\text{e}\text{c}^2}}.
\end{equation}
$\text{MTE}_\text{i}$ of 250 meV translates to $r_\beta$ of $1.4\times10^{-3}$. At the same time, $r_\text{i}$ remains a free model parameter to be found by finding the best agreement between GPT with the experiment.

The beam was launched at a charge corresponding to 20 nA and allowed to drift through the distance corresponding to a specified interelectrode gap. Here, the beam dynamics is computed self-consistently taking the COMSOL calculated field. The final distribution in $(x,y)$ and $(\beta_x,\beta_y)$ was captured at a distance corresponding to the imaging YAG screen of the microscope and are shown in Fig.\ref{F:gpt}C and D. In the $(x,y)$ space, the standard deviation $\sigma_{gpt}$ of such projections were calculated for every cathode-anode gap, and $2\sigma_{gpt}$ was taken as the computed beam diameter (analogous to FEpic image processing of the experimental images). It was established that in GPT, when $r_\text{i}$ was set to the physical radius of the fiber core of $75\,\mu\text{m}$, and $r_\beta$ was set by MTE of 250 meV,\cite{mte_fairchild} then the final diameter (spot size $2\sigma_{gpt}$) of the resulting beam projection was in a very good quantitative agreement with the experiments (see Fig.\ref{F:gapvsize}). Note, the GPT results were fairly insensitive to MTE values set between 25 (typical Fermi level value for CNT) and 250 meV, and magnification was due to the radial field distribution. This points out that the divergence between the experiment and GPT (seen for the gap ranging between 1 and 1.6 mm) stems from the difference between the idealized computed and the actual field distribution in the gap. The summary of the results in Fig.\ref{F:gapvsize} confirms that the entire whole fiber surface actively and uniformly emitting with a small $\text{MTE}$.

Finally, using Eq.\ref{E:ex_finalform} and substituting $r_\text{i}=75\,\mu\text{m}$ and $\text{MTE}_\text{i}=250\,\text{eV}$, the upper limit of the normalized emittance on the fiber cathode surface can be estimated as
\begin{equation}\label{E:num_emit}
    \epsilon_x^N=0.052\,\text{mm}\,\text{mrad} = 52\,\text{nm}.
\end{equation}
From this, taking the measured current (limited to 10--100 $\mu$A due to extremely high power density), as shown in Fig.\ref{F:condition}, the normalized brightness for 50 $\mu$A dc current is $B_N=3.7\times10^{10}\frac{\text{A}}{\text{m}^2\text{rad}^2}$. The same very fibers can draw currents of 1--10 A when operated in pulsed mode with a pulse length of 100--300 ns.\cite{fiber-IVEC2020} Using the estimated emittance of $52\,\text{nm\,rad}$, the brightness in the pulsed mode, preferable mode in most VED HPM applications, attains a notable value of $B_N=4.4\times10^{15}\frac{\text{A}}{\text{m}^2\text{rad}^2}$. This number is outstanding and is comparable with brightness metrics in the the state-of-the-art microwave/rf accelerator injectors.\cite{SOTA_Br} This (phase space) brightness can be converted into geometrical brightness, a definition of brightness commonly employed in the electron microscopy literature. The geometrical reduced brightness is defined as $B_r^G=\frac{\text{d}I}{\text{d}\Omega}\frac{1}{U}\frac{1}{S_{\text{cathode}}}$, where $\Omega$ is the solid angle, $U$ is the voltage at which the current $I$ is measured, and $S_{\text{cathode}}$ is the emission area of the cathode. Our calculations show that in pulsed mode it could attain $B_r^G=5.7\times 10^7$ A m$^{-2}$sr$^{-1}$V$^{-1}$. This number is within the range obtained for single CNT emitters.\cite{BrG_PRl2005}

\section{\label{S:conclusion}Conclusion}

In conclusion, we presented a simple and efficient field emission cathode design where CNT fiber core was plated with nickel shell. This design had two important functions. First, it compresses the core, provides mechanical strength thereby preventing stray fibril formation during conditioning and operation. Second, such design (while slightly reducing field enhancement and increasing turn-on field) reduces the fringing field on the CNT fiber and therefore the defocusing radial field.

As field emission microscopy directly demonstrated, both tested cathodes featured excellent spatially coherent emission. Field emission microscopy aided by image processing and beam dynamics simulations confirmed that the entire fiber core of 150 $\mu$m in diameter actively and uniformly emitted electrons, as well as enabled phase space analysis. All of these combined allowed to quantify the observed emission coherence through calculating emittance and brightness. The extremely low emittance resulting in record brightness highlight a simple and practical path forward for the CNT fiber technology that has long been expected to advance high frequency vacuum power devices but had limited success due to low brightness.

Finally and most importantly, it was demonstrated that the nanoscopic single CNT cathode technology can be translated to the macroscopic fiber CNT level in terms of emission uniformity. In other words, spatial coherence and uniformity (intrinsic to a single CNT emitter) can be achieved in a CNT fiber comprised out of billions of single CNT's. The obtained brightness figures of merit further confirm this technology translation in that ultimate single CNT emitter brightness is feasible to attain for CNT fiber cathodes.

\bibliography{references}

\end{document}